\documentclass[review]{elsarticle}
\usepackage{graphics}
\usepackage{graphicx}
\usepackage{float}%place images here with [H]
\usepackage{bm}
\usepackage{tabularx}
\usepackage{algorithm}
\usepackage{algpseudocode}
\usepackage[nomarkers,notablist,nofiglist]{endfloat}
\usepackage{url}

\usepackage{subfigure}

\usepackage{lineno}
\modulolinenumbers[5]
\usepackage{siunitx}         
\graphicspath{ {./Figures/} } 
\usepackage{soul}
\usepackage{miller}
\usepackage{amsmath}
\usepackage{geometry} % to change the page dimensions
\usepackage{color} 
\usepackage{outlines}
\usepackage{upgreek}
\journal{arXiv}

\begin{document}

\begin{frontmatter}

\title{An Automated Magnetron Sputtering Chamber for Ferroelectric Thin Film Deposition}

\author[1,2]{Stanislav A. Udovenko}
\author[2]{Ian Mercer}
\author[2]{Sarah Olandt}
\author[1]{Ric Wilburn}
\author[1]{Kevin Dressler}
\author[1,2]{Susan Trolier-McKinstry}
\author[2]{Jon-Paul Maria}
\author[2]{Darren C. Pagan \corref{mycorrespondingauthor}}
\ead{dcp5303@psu.edu}

\address[1]{Materials Research Institute, The Pennsylvania State University, University Park, PA 16802, United States of America}
\address[2]{Materials Science and Engineering, The Pennsylvania State University, University Park, PA 16802, United States of America}

\begin{abstract}

Optimization of next-generation materials synthesis and manufacturing processes can be accelerated by effective use of digital datasets. However, a majority of existing custom research infrastructure, including that for thin film deposition, is primarily manually operated and not compatible with this new research paradigm. Here, a template is provided for upgrading existing manual deposition chambers to enable automated and autonomous experimentation. As an example, the upgrade of an existing magnetron sputtering chamber dedicated to synthesis of wurtzite ferroelectrics is presented. Focus is placed on automation of instrumentation; system and deposition control; and synchronized and automated data collection strategies. An example use case of the system for semi-autonomous determination of process-property relationships is presented, specifically minimization of coercive field in wurtzite Al$_{1-x-y}$Sc$_x$B$_y$N thin films.

% This paper describes a fully automated thin-film deposition system based on a manually operated magnetron sputtering chamber used to optimize ferroelectric growth on wurtzite materials. Automation control software is LabVIEW-based. All chamber features (RF power supplies, mass flow controllers, pumps, shutters, sample stage, and in-situ characterization tools) are controlled via digital serial interfaces and analog voltages. The modular structure of the LabVIEW-based control software enables a rapid integration of new equipment into the system.

% The automated deposition system can grow complicated structures such as thick multilayer stacks with consistent layer growth quality and thickness, as well as structures with compositional gradients. The system is currently equipped with an optical ellipsometer and optical spectrometer for in-situ growth monitoring. After each deposition is completed, all datasets are automatically uploaded to an online database for machine-learning analysis and feedback.

% As part of the DOE Energy Frontier Research Center ``3D Ferroelectric Microelectronics Manufacturing'' (3DFeM$^2$) pilot project, a legacy ferroelectric sputtering chamber was converted into a fully automated deposition platform capable of complex, programmable thin-film growth. Results of SEM and electrical characterization of ten-layer AlScN/AlN stacks are reported.

\end{abstract}

\end{frontmatter}

%\linenumbers

\section{Introduction}

Traditional materials design and optimization have been, to a large degree, driven by trial-and-error based approaches. Bursts of progress come from the discovery of new physical phenomena, characterization methods, or processing methodologies, but a majority of time is spent generating incremental advances. These advances can be attained from relatively few experiments because of human abilities to synthesize, extrapolate, and hypothesize. However, human biases and limits to the number of composition and process variables that can be meaningfully analyzed often lead to neglect of large portions of synthesis space. In addition, traditional approaches are relatively slow to converge to \emph{optimal} material targets when nearing a material design goal \cite{arroyave2022perspective}, particularly as processing parameter spaces increase in dimensionality. A key benefit of automated processing, besides increasing reproducibility, is ease of interfacing with machine learning techniques (e.g., Bayesian optimization techniques) for efficient exploration of large parameter spaces \cite{wakabayashi2019machine,osada2020adaptive,shrivastava2024bayesian,ishiyama2024bayesian,zhang2024bayesian,trice2026machine}.

For these reasons, data-driven approaches that augment traditional material research and overcome the outlined challenges are of interest. These data-driven efforts include use of digital twins, advanced data analytics, machine learning (ML), and artificial intelligence (AI). Demonstrations in combinatorial materials science~\cite{ziatdinov2022hypothesis,liang2025real}, self-driving microscopy systems~\cite{kalinin2021automated, vasudevan2021autonomous,raghavan2024evolution}, and inorganic ceramic powder synthesis \cite{szymanski2021toward,szymanski2023autonomous} highlight the potential to accelerate discovery of new materials and optimize material properties.

An obstacle to implementing data-driven materials design and optimization in most academic research settings is that most materials synthesis and manufacturing infrastructure are not compatible with this vision. Tool operation is largely manual and much of the associated metadata is not retained. While larger organizations can acquire equipment compatible with data-driven research as necessary, broad dissemination of capabilities to the full research community requires facile, lower-cost strategies (i.e., upgrading existing systems) to harness the promise of data-driven research.

With this in mind, a methodology for the upgrade of an existing deposition chamber is presented here as an example of an agile approach to implementing automated thin film deposition. The upgrades for process automation and data-drive synthesis include implementation of central system control and data logging, algorithmic programming for deposition, and automated transfer of data to a centralized repository. As a demonstration of the compatibility with data-driven material synthesis, the system is used to minimize the coercive field in ferroelectric Al$_{1-x-y}$Sc$_x$B$_y$N thin films \cite{fichtner2019alscn,fichtner2025growth,skidmore2025sputtered}, as a function of N$_2$ gas flow, Sc target power, and B target power. Gaussian Process Regression was utilized to suggest the next films to synthesize to efficiently establish the process-property relationship.

\section{Deposition Chamber Automation Description}

\begin{figure}[h]
\centering
\includegraphics[width = 0.9\textwidth]{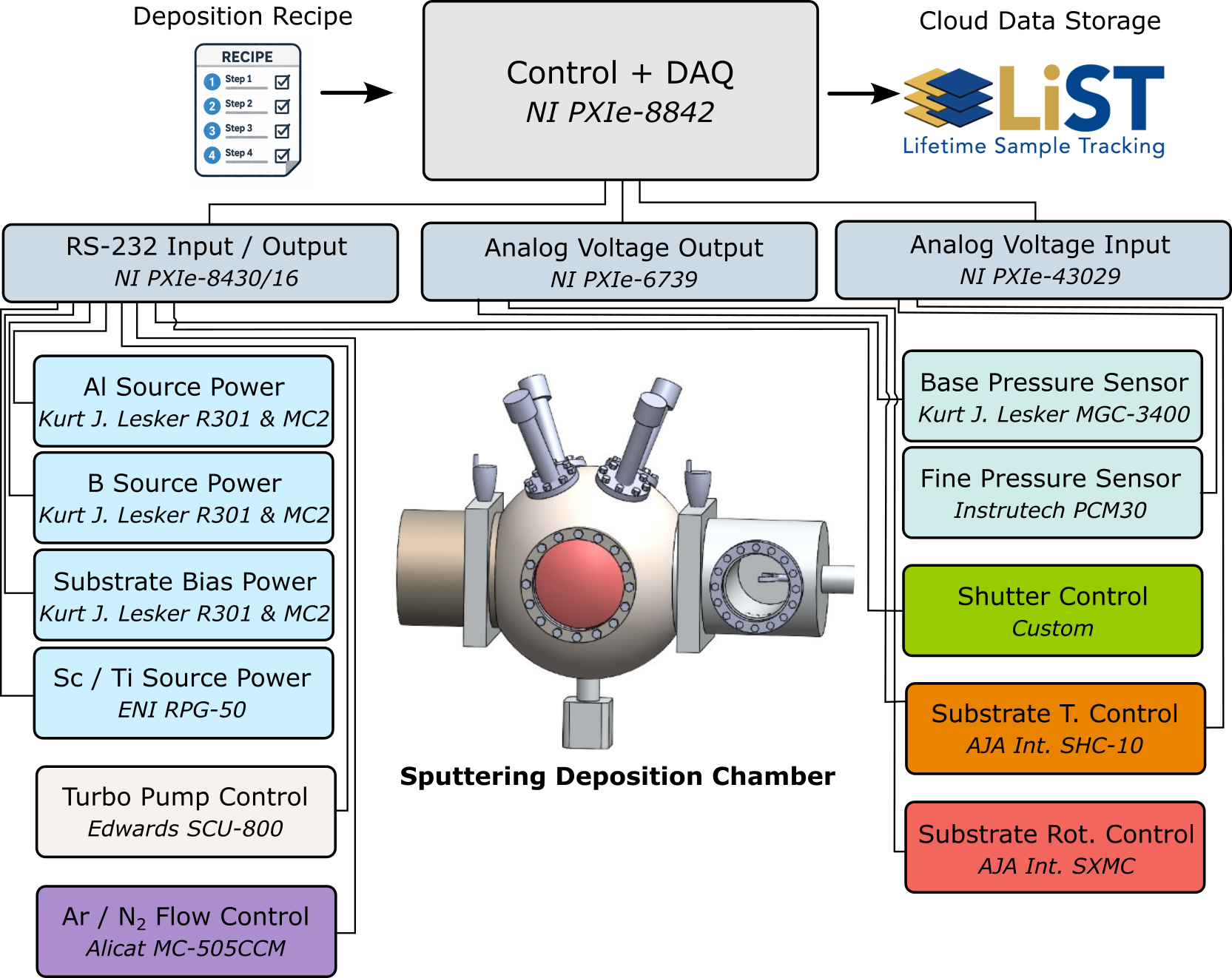}
\caption{Schematic overview of the various components and connectivity of the automated deposition chamber.}
\label{fig:schematic}
\end{figure}

The system upgrade approach utilized three primary strategies: (i) remove as much manual user input as feasible, (ii) centralize control over all deposition components, and (iii) synchronize and collate data output from disparate sensors. In addition, data are then automatically ingested into a cloud service for extended storage and accessibility. A schematic overview of the automated deposition system is provided in Figure \ref{fig:schematic}; the system is described in more detail in the following subsections.

\subsection{Hardware Retrofitting}

The existing deposition instrument was a 4-cathode confocal sputter-down tool with a 100 mm diameter substrate manipulator and RF/DC substrate bias capabilities that operated at substrate temperatures from room temperature to 900$^\circ$C. The cathodes were used for deposition of Al, B, Sc, and Ti. The first step in the process was inventorying the various components of the existing deposition chamber and assessing if/how they could be remotely interfaced. To enable automated deposition, each of the components must be addressable through an electronic connection.  Those with existing input / output (IO) serial RS-232 ports or analog voltage connections were not replaced. The existing components in the system are summarized in Table \ref{tab:components}.

In line with upgrade strategy (i), several other operations identified as manual operations were considered for automation, including sputter cathode shutter operation, substrate height adjustment, substrate shutter positioning, and substrate loading. In the existing system, the sputter cathode shutters were actuated with manual 3-way pneumatic valves. These pneumatic manual valves were replaced with pneumatic solenoid valves that could be triggered with 24 V DC signals. The existing substrate height and substrate shutter positions were adjusted by manually turned lead screws. The manual knobs were also replaced with encoded NEMA17 stepper motors to actuate motion. A custom STM32F407G-DISC1 microcontroller board was employed for valve and motor control. A RS-232-TTL converter was used to interface with the microcontroller, while a 16-channel relay module was used for valve actuation. Two CL42T closed-loop stepper motor drivers drove the sample stage and substrate shutter motors.  All of these were powered by a separate power supply. Lastly, as the growths typically take several hours and the operation was planned to be semi-autonomous (human-in-the-loop), the manual load-lock system for inserting substrates was elected to remain manual.

% An RS-232-to-UART converter translates RS-232 serial signals into microcontroller-compatible TTL-level UART signals. 
% 16-channel relay module actuates 24V pneumatic solenoid valves (source shutters).
% Two CL42T closed-loop stepper motor drivers used to drive sample stage z-axis and substate shutter motors.
% 2.8" TFT display used for real-time showing shutters and stage position.
% Power supply used for delivering power to control and executive modules.

%

\begin{table}
    \centering
    \caption{Summary of the various control units and sensors associated with sputtering deposition chamber requiring a centralized control and data acquisition.}
    \vspace{2mm}
    \begin{tabular}{|p{4.5cm}|p{0.75cm}|p{7cm}|p{1.25cm}|}
        \hline
        \textbf{Unit} & \textbf{Qty.} & \textbf{Role} & \textbf{Interface} \\
        % \hline
        % Edwards nXDS15i & 1 & Rough Pump & RS-232 \\
        \hline
        Kurt J. Lesker R301 & 3 & RF power sources for sputtering cathodes and substrate bias & RS-232\\
        \hline
        Kurt J. Lesker MC2 & 3 & Automatic RF line matching matching network controller  & RS-232\\
        \hline
        ENI RPG50 & 1 & DC pulse source to sputtering cathodes & RS-232 \\
        \hline
        Edwards SCU-800 & 1 & Turbo Pump Controller & RS-232 \\
        \hline
        Alicat MC-505CCM & 2 & Mass flow controllers Ar and N$_2$ gases & RS-232\\
        \hline
        Kurt J. Lesker MGC-3400 & 1 & Multigauge controller reading base pressure & RS-232\\
        \hline
        Instrutech PCM301 & 1 & Pressure gauge for chamber & Analog voltage \\
        \hline
        AJA Int. SHC-10 & 1 & Substrate temperature controller & Analog voltage \\
        \hline
        AJA Int. SXMC & 1 & Substrate rotation controller & Analog voltage \\
        \hline
    \end{tabular}
    \label{tab:components}
\end{table}

\begin{figure}[h]
\centering
\includegraphics[width = 0.9\textwidth]{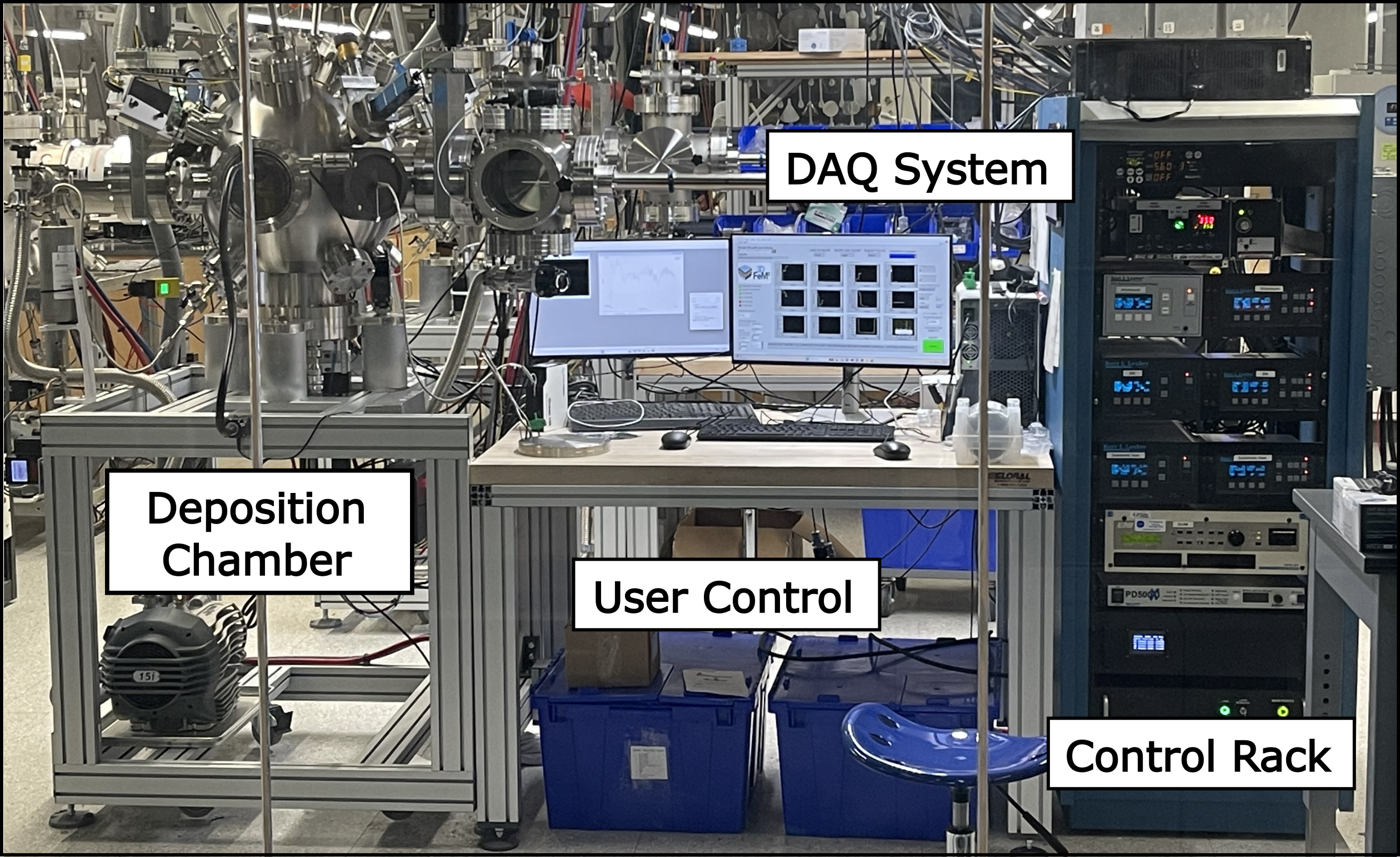}
\caption{Annotated photograph of the assembled automated magnetron sputtering deposition system.}
\label{fig:photo}
\end{figure}

A National Instruments (NI) data acquisition (DAQ) chassis solution was chosen to centrally control the electronic units described in Table \ref{tab:components} and the shutter / substrate controller. The various components of the DAQ are detailed in Table \ref{tab:NI_parts}. The primary components of the DAQ are a chassis to house a computer controller and cards for communicating with the various unit interfaces. Note that IO with RS-232 units is done with a single card, while analog units require separate input and output cards. The choice of IO cards was governed by the connections and data sampling rates necessary. Again, as growth time is on the order of hours, data collection rates in the kHz and above range were not deemed necessary. This consideration of long growth times also extends to the serial RS-232 connections for many of the units, where overhead on communications can often take on the order of seconds. A photograph of the assembled deposition system including the DAQ is shown in Figure \ref{fig:photo}.

\begin{table}
    \centering
    \caption{DAQ }
    \vspace{2mm}
    \begin{tabular}{|p{4.5cm}|p{10.0cm}|}
        \hline
        Unit & Role \\
        \hline
        PXIe-1092(VCXO) & Accommodation of controller and cards \\
        \hline
        PXIe-8842 & Run system OS, LabVIEW, and Python scripts. \\
        \hline
        PXIe-8430/16 & Communicate (IO) with units requiring RS-232 interfaces. \\
        \hline
        PXIe-6739 & Send input to units requiring analog voltage interfaces.  \\
        \hline
        PXIe-43029 & Read output from units requiring analog voltage interfaces. \\
        \hline
    \end{tabular}
    \label{tab:NI_parts}
\end{table}

% PXIe-1092(VCXO) & Accomodation of controller and cards \\
% PXIe-8842 & Run the OS, LabVIEW, Python scripts and other aux software \\
% PXIe-8430/16 & Chamber units control via RS-232 interface \\
% PXIe-6739 & Chamber units control via analog voltage  \\
% PXIe-43029 & Chamber units reading via analog voltage \\

\subsection{System Control}

\begin{figure}[h]
\centering
\includegraphics[width = 1.0\textwidth]{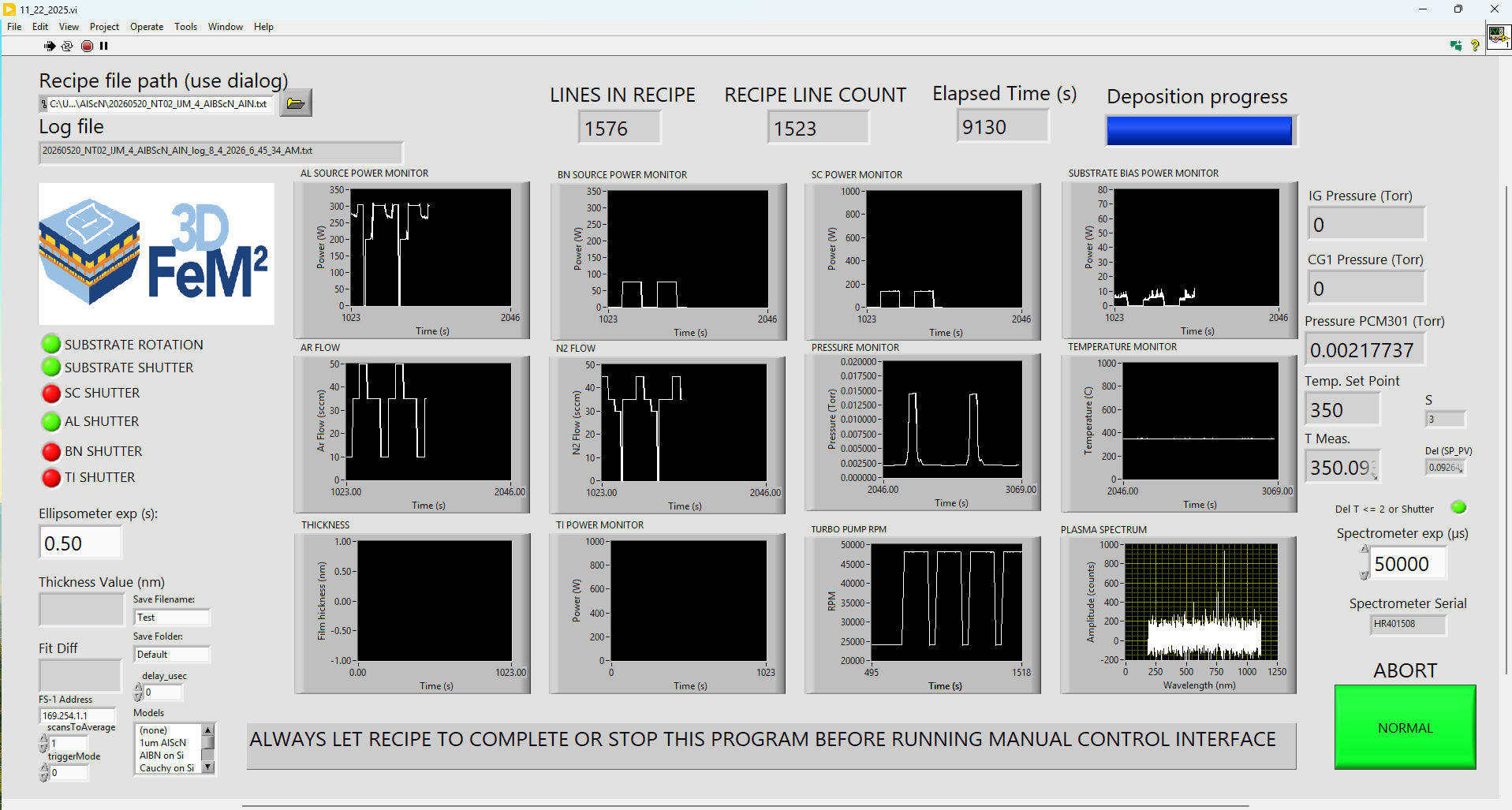}
\caption{Graphical user interface for use of the automated deposition chamber and live monitoring of the growth process.}
\label{fig:gui}
\end{figure}

To synchronize operation and data collection (upgrade strategies ii and iii), a customized control program was implemented in the NI LabVIEW programming environment. The program operates from the controller on the NI chassis and interfaces with all external units through the IO cards. The control program automates the deposition process through `recipe' files. These recipes are tables of setpoints for all electronic units used in the deposition process. Each row  of a recipe includes a collection of thermal and plasma processing steps. The number and composition of rows is defined by the user.  Two types of recipes exist that allow the system to operate in a `Time Mode' or `Event Mode.' In Time Mode, the time spent on each recipe step is fixed and defined by the user. In Event Mode, the time spent on each recipe line is defined by the system reaching user-defined unit or sensor set-points (within user-provided tolerances). For ease of use, recipes are created in human-readable spreadsheets, and a Python script converts the spreadsheets to text files for execution by the control program. Recipes can also be automatically generated by machine learning tools for semi-autonomous operation. In addition to control and data logging, the system has a graphical user interface (GUI) for user operation and to provide live views of current deposition parameters, as shown in Figure \ref{fig:gui}.

Pseudocode describing the operation of the LabVIEW controller program is given in Algorithm \ref{alg:control}. The control program begins with reading in recipe parameters and instantiating variables. User-defined parameters are then checked to be within pre-defined safe operating bounds. Next, the program loops through each recipe line, updating set points (e.g., target powers) as necessary. The program then loops through a cycle of reading in and updating unit status data (through either unique serial or analog voltage protocols for each device) followed by writing out the status data to a structured text file and the GUI. The data logging rate is user-defined but is usually collected at the rate of once per minute, sufficient for growths ranging from 30 minutes to hours. After data logging, the program checks the time or event conditions for moving to the next recipe line. If conditions have been met, the program moves to the next recipe line. The program continues until the recipe is completed or the user selects a growth abort switch on the GUI.

\subsection{Data Ingestion  and Storage}

The automated magnetron sputtering system (DOI 10.60551/rd4p-fs92) data are directly ingested into the 3DFeM$^2$ environment of the Lifetime Sample Tracking (LiST) system. LiST is a web-based data management tool that employs a custom graphical user interface and data storage system to capture experimental material sample data and the processes performed over the lifetime of those materials. Data includes but are not limited to comprehensive sample preparation information, growth recipes, synthesis protocols, and characterization data. The LiST data management system includes application programming interfaces and services that are run in a .NET framework with a React-based web application front-end and back-end services used for file and metadata ingestion.

The system data are ingested by a custom Python interface from a Microsoft Sharepoint source that supports the instrument via secure file transfer to LiST \cite{richardella2026lifetime}. Data ingestion is directed by a material sample ID key field that has a standardized character string (e.g., YYYYMMDD\_InstrumentID\_GrowerInitials\_IntradayRun\#\_Material) to differentiate unique samples. The material sample ID is linked to the sample recipe and \emph{in situ} and \emph{ex situ} characterizations (e.g., ellipsometry and X-ray diffraction respectively). The LiST system formats the user-specified key recipe data for the automated deposition system into an Microsoft SQL database table in the user interface and includes growth run specific metadata.

\section{Demonstration}

To demonstrate the capabilities of the upgraded system, a semi-autonomous process was applied to establishing a process-property relationship for Al$_{1-x-y}$Sc$_x$B$_y$N thin films. The process is referred to as semi-autonomous, as a user is still required to load and remove samples. During the process, a Bayesian optimization approach is used in which the next set of processing parameters tested are suggested in order to synthesize an optimized property (i.e., lower the coercive field) or to minimize uncertainty in the process-property relationship. The film is then grown and the property subsequently measured, completing a process iteration. The new data are then input into the Bayesian optimization process and a new thin film to test is suggested.  Three processing parameters were chosen to establish a relatively large parameter space that would be time-consuming to explore with traditional methods. The varied processing parameters used were the N$_2$ gas flow rate to adjust the N content, the power applied to the Sc target to adjust Sc content, and the power applied to the B target to adjust B content. The possible parameters were bounded between 5 and 50 sccm, 100 to 500 W, and 25 to 300 W for the N$_2$ gas flow, Sc target power, and B target power respectively.

Several processing parameters were kept fixed generally following \cite{mercer2025ferroelectric}, including the  total gas (N$_2$ and Ar) flow, substrate temperature, Al target power, substrate height, and deposition time. The fixed growth parameters are provided in Table \ref{tab:fixed_parm}. All films were grown on 4-inch, single-side-polished, highly doped $\langle$100$\rangle$ n-Si wafers (as-doped) with resistivities between 0.001-0.005 $\Omega$cm. Prior to depositions, the chamber was heated to 600$^\circ$C for 30 minutes then decreased to 350$^\circ$C for the remainder of the process. Next,  50 W of substrate bias in 40 sccm N$_2$ and 40 sccm Ar at 20 mTorr was used for 30 minutes to nitride the substrate surface. During the deposition, a substrate bias of 2 W was used. The pressure is not fixed in the chamber during depositions since different combinations of N$_2$/Ar were used but the target pressure was approximately 1.9 mTorr. The turbo pump speed varied through the recipe to reach a range of different pressures but was kept constant during each deposition step at different N$_2$/Ar flows.

\begin{table}
    \centering
    \caption{Fixed Parameters}
    \vspace{2mm}
    \begin{tabular}{|p{4.0cm}|p{2.0cm}|}
        \hline
        Parameter & Value\\
        \hline
        Substrate Bias & 2 W \\
        \hline
        Al Target Power & 300 W \\
        \hline
        Substrate Temperature & 350 $^\circ$C \\
        \hline
        Substrate Height & 50 mm \\
        \hline
        Deposition Time & 30 min.  \\
        \hline
        Total Gas Flow & 50 sccm \\
        \hline
    \end{tabular}
    \label{tab:fixed_parm}
\end{table}

Again, the goal is to learn the most likely functional relationship $f$ between the processing parameters $\bm{p}$ and the coercive field $E_C$.
\begin{equation}
f(\bm{p})=E_C
\end{equation}
To learn this relationship, Gaussian Process Regression (GPR) \cite{williams1995gaussian} is utilized. GPR samples from a normal distribution of mapping functions, $f$, (here coercive field) from the input $\bm{p}$. The prediction of the model is the mean mapping function ($\bar{f}$) from the normal function distribution. The mapping functions learned are linear combinations of input training data $\bm{p}^*$, where the coefficients are distances (usually Euclidean) between new predictions and input training data. The weights are also a function of a chosen covariance (or kernel) function $K$. Here a rational quadratic function is used \cite{duvenaud2014automatic}
\begin{equation}
K_{ij}=\left(1+\frac{||\bm{p}_i-\bm{p}_j||^2}{2\alpha}\right)^{-\alpha}
\end{equation}
where $\alpha$ is a hyperparameter (here 0.05) and the indices $i$ and $j$ indicate different measurements at different iterations. In addition, for each function prediction, a variance $\sigma^2(\bm{p})$ from the sampled functional space is also calculated, which again is a function of the distance between the new prediction and training data. Generally regions which lack training data in the nearby vicinity have high uncertainty. From current function predictions and associated uncertainty, the next set of test parameters $\bm{p}_{n+1}$ are chosen. The choice of next processing parameters follows the relationship 
\begin{equation}
\bm{p^*}_{n+1}= 
\begin{cases} \mathrm{arg~min}(\bar{f}) & \mathrm{if} \;\mathrm{min}(\bar{f}(\bm{p^*}_{n+1})) < \kappa \, \mathrm{min}(\bar{f}(\bm{p}_{n})) \\
\mathrm{arg~max} (\sigma) & \mathrm{else} \end{cases} \quad .
\label{eq:alg}
\end{equation}
where $\kappa$ is a parameter to adjust the balance between testing a lowest $E_C$ prediction and searching in regions of high uncertainty.  For this work $\kappa$ is set to 0.95. The logic is that if the minimum coercive field predicted gives a 5\% improvement from the previous step, that set of processing parameters is tested, otherwise, the coercive field prediction with the highest uncertainty is tested. With this approach the film with the lowest coercive field is generated while trying not to miss local minima in unexplored regions of parameter space.

\begin{figure}[h]
\centering
\includegraphics[width = 1.0\textwidth]{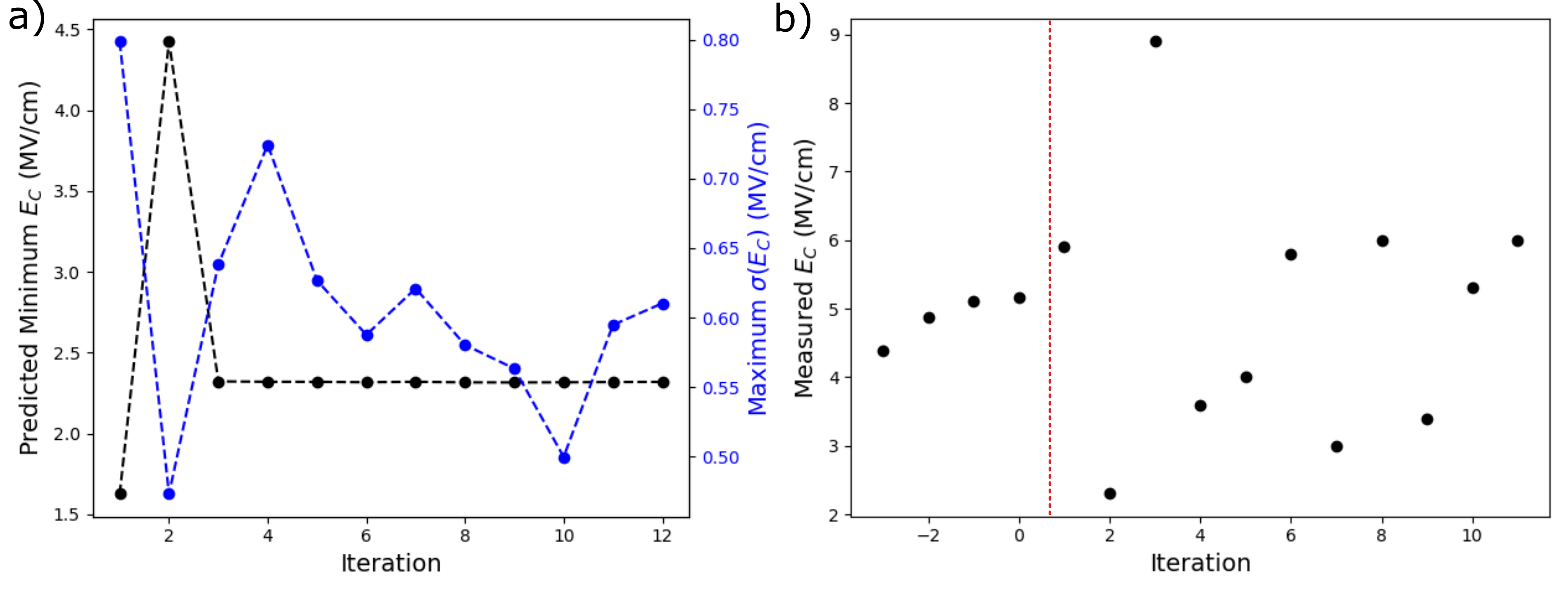}
\caption{a) Evolution of the predicted minimum coercive field $E_C$ (black) and maximum standard deviation $\sigma$ of the predictions (blue) for the Al$_{1-x-y}$Sc$_x$B$_y$N films with Gaussian Process Regression (GPR) iteration. b) Measured coercive field for the film grown at each GPR iteration. The red dashed line indicates when the GPR model began to be used to generate deposition recipes.}
\label{fig:evo}
\end{figure}

To instantiate the GPR model, four films were produced (iterations) with randomly-selected processing parameters (N$_2$ gas flow, Sc target power, and B target power)  within the processing bounds. The film heights and coercive fields were then input into the GPR model to predict the coercive field $E_C$ and the standard deviation $\sigma$ of the predictions across the processing parameter space. Ten subsequent films were produced using recipes with varied processing parameters dictated by the logic in Eq. \ref{eq:alg}. Figure \ref{fig:evo}a shows the evolution of the predicted minimum coercive field $E_C$ and the maximum standard deviation $\sigma$ (highest uncertainty) across the processing parameter space. Figure \ref{fig:evo}b shows the measured coercive field of the ten films produced at each iteration with the red dashed line indicating when GPR began to be used to guide film growth. As can be seen, the minimum $E_C$ predictions and uncertainty oscillate until Iterations 3 and 4 respectively when the predictions begin to stabilize. At Iteration 7, a film was produced with the minimum coercive field of 2.3 MV/cm and after this point the algorithm attempted to fill in the processing space. The growths were stopped at 10 films because it appeared that field minimum had been located (see below) and the maximum uncertainty was only marginally decreasing. The composition of the film with the lowest coercive field (2.3 MV/cm) was determined with X-ray photoelectron spectroscopy to be Al$_{0.668}$Sc$_{0.33}$B$_{0.002}$N.

\begin{figure}[h]
\centering
\includegraphics[width = 1.0\textwidth]{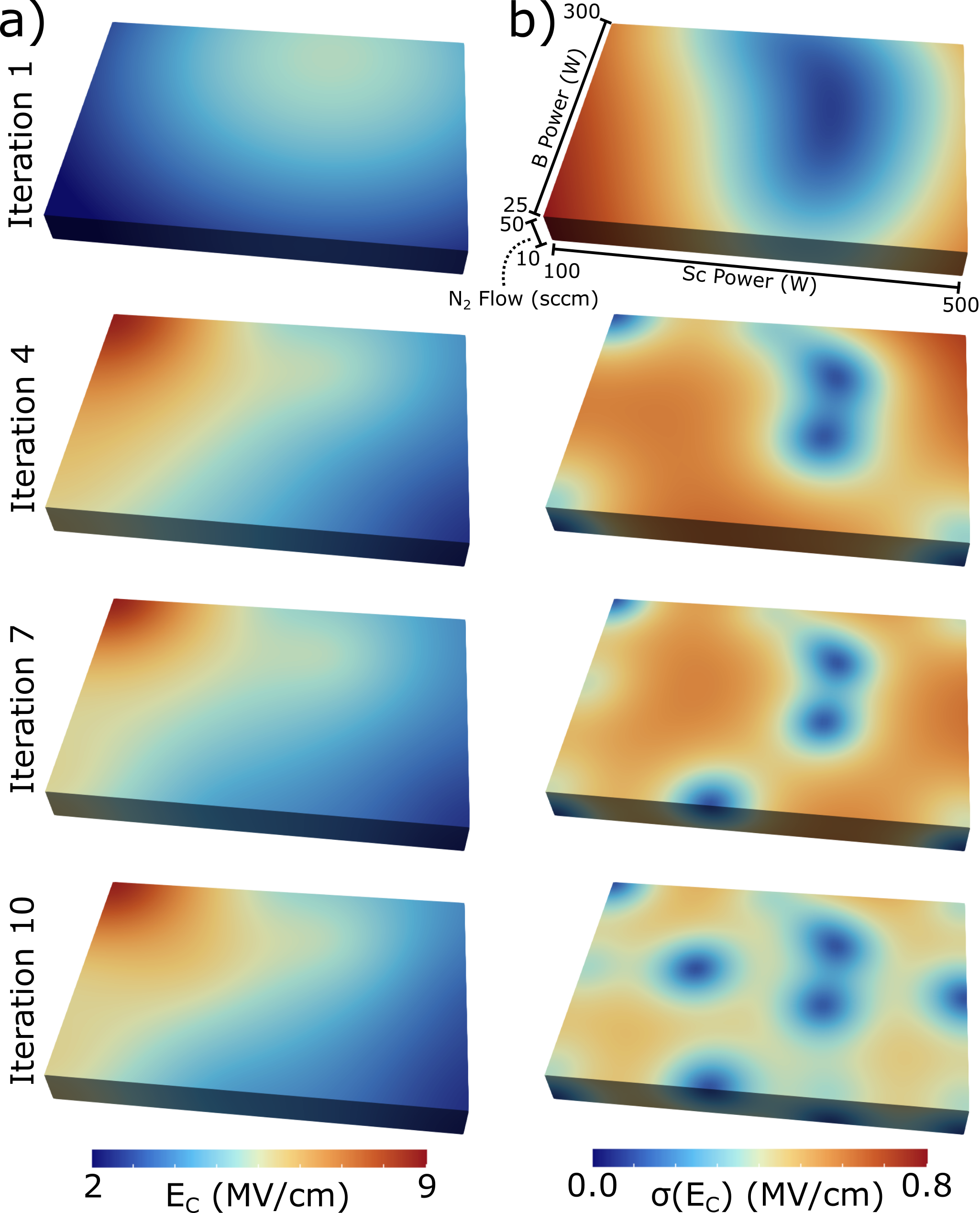}
\caption{a) Evolution of $E_C$ predicted by GPR at Iterations 1, 4, 7, and 10 for the processing space consisting of N$_2$ flow rate, Sc target power, and B target power. b) Evolution of standard deviation $\sigma$ of the GPR $E_C$ predictions at the same iteration points.}
\label{fig:fields}
\end{figure}

To more directly investigate the evolution of the field predictions, Figure \ref{fig:fields}a
shows the evolution of the GPR predictions of the full processing-property space with increasing iterations (1, 4, 7, and 10). Figure \ref{fig:fields}b shows the evolution of complementary uncertainties (standard deviations) for the $E_C$ predictions. As there are three processing parameters, the predictions span a 3D space with each dimension being labeled. Similar to Figure \ref{fig:evo}, the predictions of $E_C$ vary significantly until Iteration 4, which then remains stable as further measurements are made. As can be seen in Iteration 10, high B target power and low Sc target power produce the highest coercive fields. In addition, the results suggest that increasing Sc power (and Sc content) could further decrease the coercive field. However, the maximum Sc power was selected for system safety. The N$_2$ flow rate only modestly affected the coercive field, but flow rates near the lower bound of 10 sccm had reduced coercive fields.

\section{Summary and Conclusion}

The automation of an existing magnetron sputtering deposition chamber was presented with a focus on strategies for implementing automation on existing academic-scale tools with modest investment. The automation implemented here facilitated the use of the deposition system to minimize the coercive field in Al$_{1-x-y}$Sc$_x$B$_y$N thin films through semi-autonomous, Bayesian process parameter optimization. Efforts such as those presented here are critical to utilize data-driven research and materials discovery strategies. It is emphasized that automation efforts for existing equipment extend beyond motorization and developing instruction protocols, but also to the automation of \emph{data collection}. The collection of system status and sensor signals needs to be centralized and synchronized along with being automatically saved to accessible repositories to be most useful for future data analysis. In this work, the integration of the upgraded instrument into the LiST data collection platform was a critical part of enhancing the utility of the instrument for data-driven materials science.

\section*{Data Availability Statement}

The recipes, deposition logs, and X-ray diffraction data for the Al$_{1-x-y}$Sc$_x$B$_y$N thin films generated during the semi-autonomous optimization of coercive field are publicly available on the LiST data management system at \url{https://m4-3dfem2-list.ad.psu.edu/data/10P-ZRQw2JkTbMY9}.

\section*{Acknowledgments}

This work was supported by the Center for 3D Ferroelectric Microelectronics Manufacturing (3DFeM$^2$), an Energy Frontier Research Center funded by the U.S. Department of Energy, Office of Science, Basic Energy Sciences, under Award No.~DE-SC0021118. We thank Dr. Alexandru Marin for help performing the X-ray photoelectron spectroscopy measurements.

\newpage

\bibliographystyle{elsarticle-num}
\bibliography{ref.bib}

\clearpage
\begin{algorithm}
\caption{System control algorithm implemented in NI LabVIEW}
\begin{algorithmic}
\State $\text{timeModeFlag}\gets \text{Read User Input}$ 
\State Read Recipe File and Set [timeTargets] or [sysTargets] \Comment{Time Mode or Event Mode}
\State $\text{abortFlag}\gets \text{False}$
\State $\text{[sysVals]}\gets$ Read Sensor Values and Unit Statuses
\For{$i \gets 1$ to $\text{numRecpLines}$}
    \State Set Unit Setpoints to [sysTargets[i]]
    
    \If{$\text{timeModeFlag}=\text{True}$} \Comment{Time Mode Operation}
        \State $t\gets 0$ 
        \While{$t<\text{timeTargets[i]}$}  
            \State $\text{[sysVals]}\gets$ Read Sensor Values and Unit Statuses
            \State Write $\text{[sysVals]}$ to Text Log and Update GUI
            \State $\text{abortFlag}\gets$ Read Abort Switch
            \State $t\gets$ Read Current Time
            
            \If{$\text{abortFlag}=\text{True}$}
                \State $t\gets \text{recpTime[i]}+1$       
            \EndIf
            
        \EndWhile
    \Else \Comment{Event Mode Operation}
        \While{$|\text{[sysVals]}-\text{[sysTargets[i]]}|<\text{[tolerances]}$}  

            \State $\text{[sysVals]}\gets$ Read Sensor Values and Unit Statuses
            \State Write $\text{[sysVals]}$ to Text Log and Update GUI
            \State $\text{abortFlag}\gets$ Read Abort Switch
            \If{$\text{abortFlag}=\text{True}$} 
                \State $\text{[sysVals]} \gets \text{[sysTargets[i]]}$       
            \EndIf
            
        \EndWhile
    \EndIf

    \If{$\text{abortFlag}=\text{True}$} 
        \State $i\gets \text{numRecpLines}+1$      
    \Else
        \State $i\gets i+1$
    \EndIf

\EndFor

\end{algorithmic}
\label{alg:control}
\end{algorithm}

% \begin{algorithm}
% \caption{LabVIEW control algorithm simplified}
% \begin{algorithmic}
% \For{$i = 1$ to $\text{Number\_of\_recipe\_lines}$}
%     \While{$\lvert T \rvert < T_{\text{SET}}$}
%         \State Execute $i$-th line of recipe
%     \EndWhile
% \EndFor
% \end{algorithmic}
% \end{algorithm}

% \begin{algorithm}
% \caption{LabVIEW control algorithm extended}
% \begin{algorithmic}
% \State $flag \gets \text{true}$
% \For{$i = 1$ to $\text{Number\_of\_recipe\_lines}$}
%     \While{$\lvert T \rvert < T_{\text{SET}}$}
%         \For{$j = 1$ to $2$}
%             \If{$flag$}
%                 \State Write to devices
%                 \State $flag \gets \text{false}$
%             \Else
%                 \State Read devices
%                 \State $flag \gets \text{true}$
%             \EndIf
%         \EndFor
%     \EndWhile
% \EndFor
% \end{algorithmic}
% \end{algorithm}

\end{document}